\documentclass[reviewcopy]{elsart}

\usepackage{natbib}
\usepackage{lineno}

\usepackage{epsfig}

\usepackage{amssymb}

\newcommand{\methane}{CH$_4$}
\newcommand{\methyl}{CH$_3$}
\newcommand{\methylene}{CH$_2$}
\newcommand{\ethyl}{C$_2$H$_5$}

\newcommand{\meththirteen}{$^{13}$CH$_4$}
\newcommand{\dmeththirteen}{$^{13}$CH$_3$D}

\newcommand{\cyanoacet}{HC$_3$N}
\newcommand{\cyanogen}{C$_2$N$_2$}

\newcommand{\hydrogen}{H$_{2}$}
\newcommand{\nitrogen}{N$_{2}$}
\newcommand{\coo}{CO$_2$} 
 
\newcommand{\dmethane}{CH$_{3}$D}
\newcommand{\diacet}{C$_{4}$H$_{2}$}

\newcommand{\methacet}{C$_{3}$H$_{4}$}
\newcommand{\acet}{C$_{2}$H$_{2}$}
\newcommand{\acetthirteen}{$^{13}$C$^{12}$CH$_2$}

\newcommand{\ethane}{C$_{2}$H$_{6}$}
\newcommand{\propane}{C$_{3}$H$_{8}$}

\newcommand{\ethylene}{C$_{2}$H$_{4}$}

\newcommand{\micron}{$\mu$m}

\newcommand{\cm}{cm$^{-1}$}
\newcommand{\radunit}{${\rm W \: cm^{-2} \: sr^{-1} / cm^{-1}}$}

\newcommand{\cmm}{cm$\:$molecule$^{-1}$}

\newcommand{\apriori}{{\em a priori}}

\newcommand{\dg}{$^{\circ}$}

\def\baselinestretch{1.6}

\begin{document}
\linenumbers

\parindent 0mm
\Large

\centerline{\bf Titan's Prolific Propane: The Cassini CIRS Perspective}

\vspace*{1cm}
\large 
\centerline{C.A.~Nixon$_{a,b}$, D.E.~Jennings$_b$, J.-M.~Flaud$_c$,
  B.~B{\'e}zard$_d$, N.A.~Teanby$_e$}
\centerline{P.G.J.~Irwin$_e$, T.M.~Ansty$_f$, A.~Coustenis$_d$,
  S.~Vinatier$_d$, F.M.~Flasar$_b$} 

\vspace*{0.5cm}
\centerline{$^*$ Corresponding author: {\tt e-mail: conor.a.nixon@nasa.gov}}

\vspace*{0.5cm}
\normalsize
\centerline{$_a$University of Maryland, College Park, MD 20742, USA}
\centerline{$_b$NASA Goddard Space Flight Center, Greenbelt, MD 20771,
U.S.A.}
\centerline{$_c$CNRS, Universit{\'e}s de Paris Est and Paris 7,}
\centerline{61 Av. G{\'e}n{\'e}ral de Gaulle, 94010 Cr{\'e}teil, France}
\centerline{$_d$LESIA, Observatoire de Paris, CNRS, 5 Place Jules Janssen,
92195 Meudon Cedex, France}
\centerline{$_e$Atmospheric, Oceanic and Planetary Physics,
University of Oxford, Clarendon Laboratory,}
\centerline{Parks Road, Oxford, OX1 3PU, UK}
\centerline{$_f$Cornell University, Ithaca, NY 14853, U.S.A.}

\large
\vspace*{1cm}
\centerline{\bf Submitted to {\em Planetary and Space Science}, \today}
\vspace*{0.5cm}
\vspace*{0.5cm}
\centerline{(figures=7 tables=3)}
\newpage


\Large
Running Head: \newline
\centerline{\bf Titan's Prolific Propane}

\vspace*{3cm}
\large
Direct correspondence to: \newline
Conor A. Nixon \newline
Solar System Exploration Division \\
Planetary Systems Laboratory - Code 693 \\
NASA Goddard Space Flight Center \newline
Greenbelt \newline
MD 20771 \newline
U.S.A. \newline
tel. (301) 286-6757 \newline
fax. (301) 286-0212 \newline

\newpage

\parindent 6mm
\LARGE
\centerline{\bf ABSTRACT}
\normalsize 

Although propane gas (\propane ) was first detected in the stratosphere of
Titan by the Voyager IRIS infrared spectrometer in 1980, obtaining an
accurate measurement of its abundance has proved difficult. All
existing measurements have been made by modeling the $\nu_{26}$ band
at 748~\cm : however different analyzes over time 
have yielded quite different results, and it also suffers from
confusion with the strong nearby $\nu_5$ band of
acetylene. 
In this paper we select large spectral 
averages of data from the Cassini Composite Infrared
Spectrometer (CIRS) obtained in limb-viewing mode at low latitudes 
(30\dg S--30\dg N), greatly
increasing the path length and hence signal-to-noise ratio for
optically thin trace species such as propane. By modeling and
subtracting the emissions of other gas species, we demonstrate that at
least six infrared bands of propane are detected by CIRS, including
two not previously identified in Titan spectra. Using a
new line list for the range 1300--1400~\cm , along with an
existing GEISA list, we retrieve propane abundances from two
bands at 748 and 1376~\cm . At 748~\cm\ we retrieve
$4.2 \pm 0.5 \times 10^{-7}$ (1-$\sigma$ error) at 2 mbar, 
in good agreement with previous studies, although lack of hotbands 
in the present spectral atlas remains a problem. 
We also determine $5.7 \pm 0.8 \times 10^{-7}$ at 2 mbar from the 
1376~\cm\ band - a value that is probably affected by systematic errors
including continuum gradients due to haze and also an imperfect model
of the $\nu_6$ band of ethane. 
This study clearly shows for the first 
time the ubiquity of propane's emission bands across the
thermal infrared spectrum of Titan, and points to an urgent need for
further laboratory spectroscopy work, both to provide the line positions and
intensities needed to model these bands, and also to further
characterize haze spectral opacity. The present lack of accurate modeling
capability for propane is an impediment not only for the measurement
of propane itself, but also for the search for the
emissions of new molecules in many spectral regions. \\

\vspace*{1cm}
\centerline{\bf key words=TITAN ATMOSPHERE; ATMOSPHERIC ABUNDANCES,}
\centerline{\bf OUTER PLANETS; PROPANE; INFRARED SPECTROSCOPY;}
\centerline{\bf ABUNDANCE RETRIEVAL}

\newpage
\normalsize

\section{Introduction}
\label{intro}

Propane (\propane ) was first detected in the atmosphere of Titan by
the Voyager 1 IRIS spectrometer during the 1980 encounter
\citep{maguire81}, when excess emission at 748, 922, 1053 and 1158
\cm\ was used to make the identification. Propane is an end-product of
Titan's active photochemistry. This process begins when methane is
dissociated in the upper atmosphere by UV photons and energetic
particles to form hydrocarbon radicals and ions, which undergo a
complex chain of reactions to form larger molecules and haze
particles. The main pathway leading to  \propane\ is the three-body
association: \methyl\ +  \ethyl\ + M $\rightarrow$ \propane\ + M
\citep{yung84, wilson04}, and most is ultimately removed by condensation
in the lower stratosphere. Spectrally, propane is a very important
molecule, as it has 27 predicted infrared vibrational modes of which
at least 23 are active (see \S 4.1).

Early attempts to measure the abundance of propane from the Voyager
dataset suffered from a lack of reliable laboratory spectroscopic
data, resulting in initial over-estimates followed by a downward
trend of revised abundances until the mid-1990s. \citet{maguire81}
used laboratory spectra to compute the band intensity of the 748 \cm\
band, thereby inferring a uniformly-mixed abundance for Titan's
stratosphere of $2\times 10^{-5}$, with an uncertainty factor of
three. The value was halved the next year by \citet{kim82}, who
modeled the same band to obtain a propane abundance of $1.2\times
10^{-5}$. \citet{coustenis89a} also used a band-model approach, based
on newly measured absorptions from \citet{giver84}, and a full
radiative transfer treatment to greatly revise the estimate for
uniformly-mixed stratospheric propane downwards to $(7\pm 4)\times
10^{-7}$ at the equator.

In 1991 $\sim$9000 lines representing the 748
\cm\ band were added to the GEISA atlas \citep{husson92},
with the data taken from unpublished measurements at NASA GSFC by
G. Bjoraker circa 1986. These allowed \citet{coustenis95} to make a
full line-by-line model of the blended \acet\ and \propane\
emissions. For the first time, the latitude variation of propane
was reported: a modest increase from $5.0 \pm 1.7 \times 10^{-7}$ 
at the equator to $1.2 \pm 0.5 \times 10^{-6}$ at 70\dg N 
(assuming uniform stratospheric abundances). Simultaneously, 
enhancements in many other minor gases were detected
in the northern (winter) polar stratosphere: especially in the heavier,
acetylene-derived gases (\diacet , \methacet , \cyanoacet , \cyanogen
), but also HCN and \ethylene . These enhancements have been modeled
and explained dynamically, as due to downward advection of short-lived
species at the pole(s) in a global circulation cell(s), as these
molecules tend to have higher abundances at higher altitudes in the
absence of dynamics \citep[][and references therein]{teanby09,lebonnois09}.

The launch of the Infrared Space Observatory (ISO) in 1995 provided a
new tool for studying planetary atmospheres. In 1997, the Short
Wavelength Spectrometer (SWS) was used to acquire spectra of Titan
with a mean resolution of 0.4 \cm\ - an order of magnitude improvement 
over the 4.3 \cm\ of Voyager IRIS, although spatially unresolved. Low-latitude
gaseous abundances were reported by \citet{coustenis03}, and included
a measurement of $q_{\rm C3H8} = 2 \pm 1 \times 10^{-7}$ via the same
(748 \cm ) band as Voyager IRIS. The reason for this slightly lower
value is unclear, although real seasonal change might be a factor, as
well as the higher spectral resolution.

Even this improved spectral resolution was insufficient to actually 
separate the emission lines of propane at 748 \cm\ from those of the 
R-branch lines of the 729 \cm\ ($\nu_5$) band
of \acet , and weaker contributions from HCN and \ethane . The
first study to resolve the \propane\ emission features from the other
gases was made by \citet{roe03}, who used the R$\simeq 10^5$ TEXES
Echelle Spectrograph attached to the NASA Infrared Telescope Facility
(IRTF) to make high spectral resolution, disk averaged observations of
Titan, resulting in a stratospheric abundance estimate of $(6.2\pm
1.2)\times 10^{-7}$ at 90--250 km (13.0--0.24 mbar), 
in agreement with the low-latitude IRIS values. It is important to
note that these authors compared four line lists for the 748 \cm\
propane band then available, noting significant discrepancies between
all the lists and actual propane spectral data. Only the absorption
coefficients that they derived empirically from laboratory data
provided a good fit to the planetary data. In addition, this was the
first study of Titan's propane to use the $\nu_{26}$ designation
for this band, rather than the $\nu_{21}$ designation that was
commonly used by earlier authors. We discuss the dichotomy of
propane band designations further in section \ref{sect:bands}.

In 2004 the Cassini/Huygens spacecraft entered orbit around Saturn,
opening a new era in Titan research. The Composite Infrared
Spectrometer (CIRS) instrument onboard Cassini is a successor to the
Voyager IRIS, although with many improvements including to spectral
range and resolution, and spatial resolution (see Section \ref{sect:cirsobs}).
Furthermore, the great variety of Titan flyby encounter ranges and
relative orbital inclinations, combined with the much greater
available observing time versus the Voyager encounters, permits
Cassini CIRS to make a very diverse range of scientific
investigations. These include low-spectral resolution North--South
surface temperature mapping \citep{jennings09} and limb temperature
mapping \citep{achterberg08a}, through to
high spectral resolution composition measurements on both the disk and
the limb \citep[][and references
  therein]{coustenis07,vinatier07a,teanby08a}, 
with vertical resolution of better than one scale height (40--50~km).

So far, two analyses of CIRS Titan spectra have considered
propane. \citet{coustenis07} modeled zonally-averaged sets of
mid-infrared spectra (600--1400 \cm ) from 70\dg S--70\dg N to measure
the latitude variations in hydrocarbons, nitriles and \coo , including
the 748 \cm\ band of propane for which spectral line data is available
in GEISA. \propane\ was found to exhibit a very slight increase from
the south to the north, from $\sim 5\times 10^{-7}$ to $\sim 8\times
10^{-7}$.  \citet{vinatier07a} modeled two high spectral resolution
(0.5 \cm ) observation sets of Titan's limb, from the Tb and T3 flybys
and directed at 15\dg S and 80\dg N respectively. \propane\ was found
to display an increasing abundance with altitude in the range
$\sim$150--350~km ($\sim$ 2.0--0.01 mbar) 
at both latitudes, in accordance with the
conventional understanding of photochemical formation in the upper
atmosphere and downward advection to the lower stratosphere where
removal through condensation occurs \citep[89\% according to][]{wilson04}.

In this paper we have two objectives. First, we show clearly
the very prolific extent of the multiple propane bands that can be
detected by CIRS across the mid-infrared. At least six bands are visible
in limb spectra after the modeling and removal of other
stronger gas emissions. Second, we use a new laboratory line list
for propane in the range 1300--1400~\cm\ \citep{flaud01}, along
with the GEISA list for the 748 \cm\ band to make independent
measurements of the propane abundance, and show that the results
are probably compatible when systematic errors are accounted for.
Finally, we draw conclusions about the current status of propane
retrievals in the thermal infrared.

\section{Data Acquisition}

In this paper we make use of two spectral datasets. The first is a
set of Titan limb spectral data, acquired by Cassini CIRS between 2004 and
2008. The second is a set of laboratory absorption spectra of room
temperature propane gas acquired at Pacific Northwest National
Laboratory in 2002. Both datasets are now described.

\subsection{Cassini CIRS Limb Observations}
\label{sect:cirsobs}

The Composite Infrared Spectrometer (CIRS) carried on-board Cassini is
a dual Fourier Transform interferometer design \citep{kunde96, flasar04b}. A
shared telescope and foreoptics feed incoming radiation simultaneously
into two interferometers: a Martin-Puplett (polarizing) type
optimized for far-infrared rays (10--600 \cm ), and a conventional
Michelson design for the mid-infrared (600--1400 \cm ). The recombined
beam of the far-infrared interferometer is focused onto a single
circular bolometer detector, known as focal plane 1 (FP1), while
mid-infrared detection capability is provided by two $1\times 10$
HgCdTe arrays: focal plane 3 (FP3, 600--1100 \cm ) and focal plane 4
(FP4, nominal range 1100--1400 \cm , see also Appendix A). 
Both interferometers also share a common scan
mirror mechanism and reference laser. The spectral resolution is
determined by the commanded scan time, and is variable from 15.5 to
0.5~\cm .

At 5--9~hours from closest approach to Titan (1.0--1.8$\times 10^5$
km range), the CIRS instrument makes observations of the atmospheric
limb, placing the mid-infrared arrays perpendicular to the disk edge,
and obtaining spectral samples at $\sim$30--50~km resolution,
equivalent to approximately one atmospheric scale height. These
observations come in two types: (i) temperature maps (shorter dwell,
15.5~\cm\  spectral resolution, repositioning at multiple latitudes)
and (ii) composition integrations (longer dwell, 0.5~\cm\ resolution,
single latitude).

For the purpose of identifying the weak emission bands, we have created
large averages of low-latitude spectra for FP3 and FP4 from the high
spectral resolution limb integrations (type (ii) above). 
Note that we do not attempt to
explore the vertical and latitudinal variation of the propane
abundance from the CIRS dataset, which has been documented elsewhere
\citep{coustenis07, vinatier07a, vinatier09}, but rather to maximise 
signal-to-noise to emphasize and detect very weak emission features
while still maintaining reasonable homogeneity of the dataset.

Therefore we chose to average across all flybys from the prime mission (June
2004 to June 2008) and latitudes in the range 30\dg S--30\dg
N, in the knowledge that Titan's stratosphere is slowly varying in this 
temporal and spatial interval \citep{coustenis07,
achterberg08a, teanby08b}. However, Titan's atmosphere varies much
more rapidly in the vertical direction, and therefore we restricted our
averages to 100--150~km (pixel centers) where the signal-to-noise ratio
was found to be the most favorable. The spacecraft distance range was also
constrained to be less than $2\times 10^{5}$~km. See Table I for
additional details. 

{\bf [TABLE I]}

\subsection{PNNL Laboratory Spectroscopy}
\label{sect:pnnl}

For comparison with the CIRS spectra, Dr Steven Sharpe at
Pacific Northwest National Laboratories (PNNL), has shared with us
room temperature (296 K) absorption spectra of propane
gas, taken as part of a large-scale survey of molecular infrared
spectra. These consist of 6 individual co-added spectra of 98\% pure
propane gas measured with a Bruker-66V FTIR at 0.112 \cm\ unapodized 
resolution, covering the FP3 and FP4 part of the CIRS spectral range
(600--1400 \cm ). Full experimental details have been published in the
literature \citep{sharpe04}. 

We have convolved the PNNL spectra with a Hamming kernel
$h(\Delta\tilde{\nu})$ (instrumental
line shape) of FWHM 0.48~\cm\ for approximate comparison with the CIRS
spectra, which originate at much lower temperatures ($\sim 160$~K) 
in Titan's stratosphere:

\begin{equation}
h(\Delta\tilde{\nu}) = \frac{a \left( 1.08 - 0.64 a^2 
\left[ \Delta\tilde{\nu} \right] ^2 \right) {\rm sinc} \left(
2\pi a\Delta\tilde{\nu} \right)}
{1-4a^2 \left[ \Delta\tilde{\nu} \right]^2}
\end{equation}

\noindent
where $h$ is the weighting factor at each position
$\Delta\tilde{\nu} = (\tilde{\nu} - \tilde{\nu}_0)$ relative to a given 
wavenumber $\tilde{\nu}_0$ in the original spectrum, and $a$ is an
effective path difference: $a = 1.815 / (2 \times 0.48 ) = 1.89$~cm that
compensates the Nyquist path difference at the desired resolution
($1/(2 \times 0.48)$ cm) for the FWHM of the Hamming function in Nyquist
units (1.815). Note that $h$ is normalized so that: $\int
h (\Delta\tilde{\nu}) \: d(\Delta\tilde{\nu}) 
= 1.0$ over the desired interval: we used $\pm 10$
samples relative to each wavenumber (21-pt kernel). 

\section{Data Analysis}
\label{method}

\subsection{Model Atmosphere}
\label{sect:atm}

Our model atmosphere consists of 100 layers equally spaced in
log($p$) from the surface (1.45 bar) to 700~km ($4\times 10^{-8}$
bar.) The initial temperature-pressure profile is a smoothed version
of the Huygens HASI results \citep{fulchignoni05} (covering from the
surface to $7 \times 10^{-12}$~bar at 10\dg S, 168\dg E) 
with heights calculated from
hydrostatic equilibrium. All gases known to be relevant for the
700--1400~\cm\ range of Titan's infrared spectrum are included,
namely: \nitrogen , 
\methane , \hydrogen , HCN, \acet , \ethane , \ethylene\ and \propane
. In addition, the isotopologues \acetthirteen , \meththirteen ,
\dmethane\ and \dmeththirteen\ are included as separate species. The
initial abundances and isotopic ratios mostly follow the Huygens GCMS
results \citep{niemann05} and previous measurements by CIRS pertaining
to low latitudes \citep{coustenis07}. In particular,
\hydrogen\ and \ethylene\ are uniformly mixed, while \methane\ follows
the GCMS profile, \nitrogen\ is defined as $q_{\rm N2}=1.0-q_{CH4}$,
and the other gas primary isotopologues, with the exception of \acet ,
have constant abundances in the middle atmosphere, and follow
condensation profiles towards the tropopause. The \acet\ VMR, which is
known to have a gradient at low-latitudes, follows the 10\dg S profile of
\citet{vinatier07a}. Finally, the minor isotopologue species have
initial profiles scaled from the primary isotopologue by molecular
isotopic ratios previously determined by CIRS
\citep{bezard07,nixon08a}. Fig.~\ref{fig:gas_apriori} shows the
initial gas profiles.

\begin{figure}[h]
\caption{Appears Here.}
\label{fig:gas_apriori}
\end{figure}

Finally, our model atmosphere also includes a single aerosol component
with constant mass mixing ratio of particles above the tropopause
(45~km) as described in \citet{nixon98}.

\subsection{Forward Spectral Model}
\label{sect:fm}

The forward model \citep[NEMESIS,][]{irwin08}
computes the emerging radiation field for
Titan limb rays using the correlated-$k$ approximation
\citep{lacis91} for rapid calculation of atmospheric opacity. The
three types of opacity included are: gas vibrational bands,
collision-induced absorption (CIA) of \nitrogen , \methane\ and
\hydrogen , and aerosol/haze opacity. The $k$-coefficients for
gaseous bands and CIA opacities are computed and 
pretabulated over the full range of pressures and
temperatures required for the Titan model atmosphere, as well as the
CIRS full spectral range, to enable rapid calculations. 
The spectral opacities are pre-convolved with a Hamming apodization
function of FWHM 0.48~\cm , and 50 $g$-ordinates are used to
approximate the $k$-distributions in each spectral bin. 

Spectral line data are taken from the HITRAN 2004 
\citep[for \methane\ $\nu_4$ 1304 \cm , HCN $\nu_2$ 712 \cm\ and
  \ethylene\ $\nu_7$ 949 \cm ;][]{rothman05} and GEISA 2003
\citep[for \acet\ $\nu_5$ 729 \cm , \propane\ $\nu_{26}$ band;][]{jacquinet05}
atlases. We also used supplemental sources of line data for the 
\ethane\ $\nu_9$ band \citep[822 \cm ][]{auwera07}, for \dmeththirteen\
\citep[$\nu_6$ 1156 \cm ][]{bezard07} and for the 1300--1400~\cm\ 
bands of \propane\ \citep[mostly due to the $\nu_{18}$ 1376 \cm\ and
  the $\nu_{19}$ 1338 \cm ][]{flaud01} that post-date the most
recently available GEISA and HITRAN atlases. We made additional
scaling corrections to the line strengths for the $\nu_{26}$ band of
\propane , and also to the line width and temperature dependence
parameters for both the GEISA and the \citet{flaud01} line lists,
justified in detail in Appendix B.

Additionally, for the $\nu_6$ (1378 \cm ) and $\nu_8$ (1468 \cm ) bands of
\ethane , which are not included in any current line atlases, we used
a `pseudo-linelist' made publicly available by the Jet Propulsion
Laboratory (http:// mark4sun.jpl.nasa.gov/pseudo.html). This list is an
approximation  which seeks to model the overall absorption of these
bands by using a uniform distribution of `pseudo' lines of varying
strengths, without actual knowledge of the real line positions or
intensities.  
  
To fit the continuum we used two further sources of opacity; (i)
collision-induced opacity of six molecular pairs of \hydrogen ,
\nitrogen\ and \methane\ \citep[see detailed references
  in][]{teanby06}, and (ii) a simple haze with a constant (scaleable)
number of particles per gram of atmosphere above the tropopause. The haze
had either (A) a uniform grey spectral absorption, or (B) a spectral
absorption cross-section derived by applying Mie theory to 0.2
\micron\ radius spheres (no scattering), using the real and imaginary
refractive index data for laboratory haze analogue (tholin) as
measured by \citet{khare84}.

Previously, most models of CIRS Titan mid-infrared limb spectra 
\citep{teanby07a, vinatier07a, vinatier07b, nixon08b, jennings08,
  coustenis08a}
have used the approximation of a single limb ray originating from the
center of the detector fields of view (FOVs) projected onto the limb:
the 'infinitesimal' FOV approximation. However, \citet{teanby07b} have
investigated the spatial smoothing effect of the small but finite
mid-IR detector FOVs when applied to Titan limb sounding. They
conclude that approximating a limb view using an infinitesimal ray
introduces non-negligible errors when the highest spectral
resolution of CIRS (0.5~\cm ) is used and many spectra are co-added to
lower the NESR.

More recently \citet{nixon09} have analyzed laboratory and in-flight
field-of-view response measurements of CIRS, and computed real 1-D
spatial response functions for all 20 mid-infrared detectors. Using
these functions, they investigated systematic errors introduced by
either the infinitesimal ray approximation or the finite (or 'boxcar')
approximation for Titan limb viewing, relative to using the real
detector spatial responses. Their conclusion was that, for limb
averages of as few as 20 spectra, the errors due to both
approximations became significant compared to the noise
level. 

Therefore, in this work we have included the actual 1-D spatial
responses for each detector as follows. For the $n^{th}$ individual
spectrum, recorded by detector $m$, 
let $a_n$ be the tangent altitude (in km) at the FOV center
projected onto Titan's limb and $d_n$ be the spacecraft-tangent point
distance in km. Then the $n^{th}$ spatial weighting function quantized onto
a vertical grid $z_i$ of 1-km altitude increments is then
\citep[c.f. Eq. 10 of][]{nixon09}:

\begin{eqnarray}
w_n(z_i) & = & \frac{\overline{R}_m((z_i-a_n)/d_n)}
{\sum_{i} \overline{R}_m((z_i-a_n)/d_n)}
\end{eqnarray}.

\noindent
Where $\overline{R}_m$ is the 1-D ($Y$-averaged) response function of
detector $m$ expressed in angular units \cite[tabulated in 
Table 4 of][]{nixon09}. For the average of all $N$ spectra then the
spatial weighting function is:

\begin{equation}
W(z_i) = \frac{1}{N} \sum_{n=1}^{N} w_n(z_i)
\end{equation}.

\noindent
This `effective' spatial weighting function is used 
to compute a mean model spectrum, by weighting synthetic spectra
calculated for limb rays at each altitude: 

\begin{equation}
\bar{I}(\tilde{\nu}) = \sum_{z=z_{\rm min}}^{z=z_{\rm max}}
W(z_i) I(z_i,\tilde{\nu}) 
\end{equation}

\noindent
where $z_{\rm min}$ and $z_{\rm max}$ are the lowest and highest
altitudes where $W(z_i) \neq 0$, and $\tilde{\nu}$ is wavenumber.

In practice, the 1~km resolution calculation grid is both
unnecessarily fine to achieve sufficient modelling accuracy, and also
prohibitively slow, as it necessitates the calculation of $\sim 50$
individual limb rays to produce a single spectrum at each
modelling iteration. We performed trial retrievals at various grid
spacings, finding that 5~km grid point spacing is sufficient for
$>99$\% accuracy on both FP3 and FP4 (at which level the accuracy
becomes dominated by other effects, such as knowledge of limb
pointing), and we therefore interpolated the
function $W(z_i)$ at 5~km increments relative to the mean limb tangent
altitude of the average spectrum, cutting off at a minimum weighting
of 0.001. 5~km resolution is approximately 10\% of the scale height in
Titan's lower stratosphere.

The resampled $W(z_i)$ functions in 5~km steps for the FP3 and FP4
spectral averages analyzed herein are shown in Fig. \ref{fig:limb_convs}.

\begin{figure}[h]
\caption{Appears Here.}
\label{fig:limb_convs}
\end{figure}

\subsection{Retrieval Algorithm}
\label{sect:ret}

The retrieval algorithm is a mathematical formalism used to find the
`optimal estimate' set of model parameters, by iteration 
\citep{rodgers00}. At each step of the
retrieval, the forward model is used to calculate a synthetic
spectrum, as described above, which is then compared to the data
spectrum. A cost function is calculated which is similar to a $\chi^2$
test of the difference between the data and model spectrum, with an
additional term to provide smoothing of the solution, also known as
the a priori constraint. Model parameters are then adjusted along the
downhill gradients that tend to decrease the cost function, so that
divergence between the model and data will be reduced on the next
step. When a preset convergence limit is reached - i.e. when the
percentage change in the cost function falls below a threshold value,
the retrieval ceases. This algorithm is identical to that used in
\citet{nixon08a} - see also \citet{irwin08} for further details.

\section{Results}
\label{sect:results}

\begin{figure}[h]
\caption{Appears Here.}
\label{fig:ret_temps}
\end{figure}

In this study we first used the 1225--1325~\cm\ spectral region of the
$\nu_4$ methane band to retrieve atmospheric temperatures, holding the
methane abundance constant. Fig.~\ref{fig:ret_temps} shows the
result. Note that the retrieved temperatures in the main information
region of the lower stratosphere are significantly colder than the
HASI profile, by as much as 10~K - much greater than expected due to
latitude variation alone (HASI $=10$\dg S; CIRS $=2$\dg N). This
discrepancy has been noted and remarked on in several previous studies
of CIRS data \citep[see \S2 of][and references therein]{flasar09}, and
can be seen to be part of a larger difference where the HASI stratopause is
considerably lower (by $\sim 60$km) than that determined by fitting CIRS
data. The spectroscopic parameters for the methane $\nu_4$ band are
well-studied, and the CIRS data cannot be reconciled with the HASI
temperature profile, so a solution to this problem has not yet been
reached.

The error on the temperature profile includes both spectral noise and
also the 5\% uncertainty on the Huygens GCMS methane abundance quoted by
\citet{niemann05}. See \citet{nixon08a} for description of the error
propagation. Fig. \ref{fig:terr} shows the vertical error variation.

\begin{figure}[h]
\caption{Appears Here.}
\label{fig:terr}
\end{figure}

The temperature profile was then fixed, and we carried out
numerous separate retrievals on specific spectral regions suspected to
harbor propane emission bands. In these retrievals, the grey `aerosol'
opacity was scaled to obtain a fit to the continuum level, and
simultaneously relevant minor gas VMR profiles for each spectral
region were adjusted to fit the emission bands. 

Two separate models were calculated for each spectral region:
(i) a best fit spectrum without propane, to reveal the propane signature
in the residual, and (ii) for spectral regions where \propane\ line data
were available, we fitted a model for all gases including propane, to
obtain a \propane\ abundance, allowing the values from multiple
\propane\ bands to be intercompared. The results are now discussed in detail.  

\subsection{Propane Band Detections}
\label{sect:bands}

Figs. \ref{fig:fp3_propane} and \ref{fig:fp4_propane} show the results
of modeling various CIRS spectral regions that were candidates for
propane emissions. In each figure, the left hand column shows the data
spectrum and model (grey haze used), while the right hand column shows
the residual after subtraction of model from data, along with the corresponding
portion of the PNNL laboratory absorption spectrum. The laboratory
spectrum has been convolved with a 0.48 \cm\ Hamming kernel for direct
comparison to CIRS data, and arbitrary scaling is used.

\begin{figure}[h]
\caption{Appears Here.}
\label{fig:fp3_propane}
\end{figure}

\begin{figure}[h]
\caption{Appears Here.}
\label{fig:fp4_propane}
\end{figure}

Four strong bands are clearly detected on FP3 and four on FP4, 
as indicated on the figures, along with the locations of several
weaker undetected bands. Note
that the band at 1053~\cm\ is detected on both FP3 and FP4 where the
spectral ranges overlap. The band assignments for all
vibrational modes of \propane\ are listed in Table II.
At this juncture, we are obliged to discuss a confusion
regarding propane band numbering in the pre-existing literature.

{\bf [TABLE II]}

Historically, there has been some inconsistency
over the actual band designations for
propane's infrared spectrum. This appears to be traceable to two
original studies - a 1965 paper by Gayles and King, \nocite{gayles65}
hereafter GK65, and a 1972 paper by Shimanouchi\nocite{shimanouchi72},
hereafter S72 - that introduced differing numbering schemes (see Table II). 
The GK65 scheme was apparently followed by
\citet{maguire81}, \citet{giver84} and all subsequent analyses of the
IRIS dataset, and thence propagated into the Cassini CIRS
community. The S72 scheme however was adopted by \citet{flaud01},
\citet{roe03} and also by the U.S. National Institute of Standards
and Technology (NIST), and appears preferable now both because of that
official endorsement, and also because all 27 predicted vibrational
modes are accounted for. Therefore, we use the S72/NIST numbering scheme
in this work.  Table II lists both sets of band
designations as an aid to translating band numbers from earlier works
(especially Voyager papers).

Under the NIST numbering scheme, we clearly detect the $\nu_{26}$ (748
\cm ), $\nu_{21}$ (922 \cm ) and $\nu_{20}$ (1053 \cm ) bands on FP3,
which all have strong central Q-branch peaks. All three of these bands
were claimed to be detected by Voyager \citep{maguire81}. Note that
the shape of the $\nu_{21}$ band appears somewhat different at room
temperature (PNNL spectrum) compared to the colder Titan spectrum, due
to the presence of hot bands on the short-wavelength side of the
fundamental in the lab spectrum. The $\nu_8$ band on the other hand
lacks a Q-branch, and the detection is uncertain at this time,
although we do appear to see a minimum in the right location near
the band center at 869~\cm . Note that the residual of the
ethane fitting appears surprisingly large, given that we are using
very recently updated line parameters for the $\nu_9$ band
\citep{auwera07}. This appears to be caused by trying to minimize the
overall $\chi^2$ fit, in the presence of the non-flat continuum. 

We must be careful here to emphasize possible aerosol effects: a non-grey
aerosol could easily account for some or all of the gradient in the
residual. So long as we cannot positively distinguish between the two
opacity sources, the propane $\nu_8$ detection should be regarded as
still somewhat tentative. A clear way to make progress would be to
derive spectroscopic parameters for the band from lab data and include
it in the gas model: any remaining residual gradient would then likely
be due to aerosol alone.

On FP4, the strongest propane feature inside the clean spectral range
(1100--1400 \cm )
is the Q-branch of $\nu_{18}$ at 1376 \cm . The $\nu_7$ band
(1158 \cm ) is similar in shape to the $\nu_8$ of FP3, and is also evident
here after the modeling/removal of \dmethane\ and \dmeththirteen\
emissions. In previous works, this band has sometimes
been modeled with a band opacity model to improve the D/H
determination \citep[e.g.][]{coustenis89b}.
Other weaker Q-branch peaks of $\nu_{25}$ (1192 \cm ) and
$\nu_{19}$ (1338 \cm ) fall inside the FP4 nominal  range, but are at
the noise level. 

Proceeding with caution outside the nominal range (see Appendix A), we
see a clear signal of the Q-branch of $\nu_{20}$ (1053~\cm ), with improved
signal-to-noise (S/N) over the FP3 detection.(We can discount aliasing
here, as there is no noticeable emission at the aliased frequency of
987 \cm .)  In the short-wavelength non-clean band, we also apparently
detect the Q-branch peak of $\nu_{24}$ (1472 \cm ), although it is
apparent here that significant residual remains after modeling which
is not attributable to propane. There may be several factors at work
here to undermine our model: (i) the ethane pseudo-linelist is
likely to be imperfect; (ii) the ethane $\nu_8$ (1468 \cm ) and
$\nu_{11}$ (1469 \cm ) bands remain strong out to $\sim 1600$~\cm ,
and therefore will contribute significantly through aliasing (`folding'
around 1529~\cm ); (iii) the continuum may have haze opacity that is
difficult to assess in the presence of the other uncertainties.

\subsection{Propane Gas Abundance}
\label{sect:abund}

Having demonstrated the detection of multiple propane bands by
CIRS, we wished to model those for which line data is
available. As stated earlier, only the $\nu_{26}$ band at 748 \cm\
has been previously analyzed to obtain abundances of \propane\ on
Titan, due to its unique availability in an existing atlas (GEISA). 
However, due to more recent work by \citet{flaud01}, line parameters
for propane bands in range 1300-1500~\cm\ were available to us,
allowing the new possibility of retrieving propane abundances from
these bands also. After some experimentation, we concluded that
accurate modeling of the $\nu_{24}$ band at 1472~\cm\ was not
possible, due to the uncertainties in ethane and haze opacity,
combined with the complexitity of folding the (unknown) aliased
spectrum above 1529~\cm . However, fitting the $\nu_{18}$ at 1376~\cm\
proved to be a more tractable problem.

We used two spectral windows for the abundance measurements: 740--760~\cm\
($\nu_{26}$) and 1370--1390~\cm\ ($\nu_{18}$), choosing relatively
narrow spectral regions to avoid the effects of, respectively, the
very strong acetylene band at 729~\cm\ and large instrumental noise
spike at 768~\cm\ (FP3), and a non-flat continuum that showed apparent
broad haze features at the scale of $\sim$20~\cm\ (FP4) . Haze spectral
features are the intended subject of a future study, and are not
pursued further herein.

We then added propane gas back into the model atmosphere and
separately retrieved its best-fit abundance for each window using
the previously determined temperature profile. Our objective was not to
determine the absolute abundance of propane at a single altitude to
high accuracy, where we could have fitted all regions simultaneously,
but rather to compare the relative results for FP3 and FP4. Early
retrieval attempts using these windows alerted us to a problem with
the GEISA line strengths for the $\nu_{26}$ band, which is decribed in
detail in Appendix B. This problem was remedied by rescaling the
spectral lines prior to the final retrievals.

\begin{figure}[h]
\caption{Appears Here.}
\label{fig:prop_rets_rescale}
\end{figure}

The final results of the fitting are shown in
Fig. \ref{fig:prop_rets_rescale}. The red line shows the 
residual for a forward-model calculation using
the retrieved abundances of all gases except propane, which is removed
to show its contribution. The other lines show the residual
after fitting with all gases including propane and either grey haze
(blue line) or tholin haze (green line). The fits with the two haze
models are indistinguishable over the narrow 20~\cm\ window.

{\bf [TABLE III]}

The individual abundances determined from each model fit are given in
Table III, along with errors. Included in the error calculation (added
in quadrature) are
(i) the 1-$\sigma$ random noise on the spectrum; (ii) propagated
temperature retrieval error from two sources, (a) the spectral noise
from the temperature retrieval, and (b) the 5\% uncertainty in the
methane abundance quoted by \citet{niemann05}, combined as described
in \S~3.5 of \citet{nixon08a}; (iii) the uncertainty in the line
intensities of propane in the spectral line lists, estimated at $\sim$10\%. 

The results show that the $\nu_{18}$ yields systematically higher
values than the $\nu_{26}$, compatible at the 2-$\sigma$ error level
but not at 1-$\sigma$. We suggest that two remaining systematic
effects are the sources of the difference: (i) any residual structure 
in the continuum at a scale of less than 20~\cm\ for either spectral
region that is not included in our slowly-varying haze models, and (ii)
any imperfections in the pseudo-linelist for the the ethane $\nu_6$
band at 1378~\cm . An actual linelist for the $\nu_6$ and $\nu_8$
bands would be very valuable in eliminating this source of error. 

We also note that the $\nu_{26}$ result is in very good
agreement with the CIRS limb measurement of \citet{vinatier07a} for
15\dg S, 3 mbar: $\sim 4\pm 1.5\times 10^{-7}$, also derived from the
$\nu_{26}$ band using a different (line-by-line) model and retrieval
algorithm. Note that while \citet{vinatier07a} did not scale the GEISA
line intensities by the 0.42 factor used by us, they partly
compensated for the known problem with the $\nu_{26}$ band 
intensity by omitting the vibrational part of the partition function
when calculating the line strength. This would tend to underestimate
the line strengths, providing a similar correction factor to ours, and
the agreement between the results is not anomalous. 

\section{Summary and Conclusions}
\label{conclusions}

We have modeled a low-latitude limb spectral average of Titan,
removing all gas emissions except propane. By examining the residual,
we demonstrate that CIRS has detected at least six separate bands of
\propane, including the well-known 748 \cm\ band ($\nu_{26}$). 
We also show clearly the structure of the $\nu_7$ band at 1158 
\cm\ that underlies the $\nu_6$ band of \dmethane ; the $\nu_{21}$ and
$\nu_{20}$ bands at 922 and 1053 that were weakly detected by IRIS;
and at least two others in the range 1300--1500~\cm\
($\nu_{18}$ and $\nu_{24}$) that are evident after the removal of
methane emission. Detection of the weak $\nu_8$ centered on 869 \cm\
that underlies the wing of the \ethane\ $\nu_9$ band is uncertain.

We have modeled the $\nu_{26}$ band using GEISA line data, and also
the $\nu_{18}$ band using spectral line data
from the work of \citet{flaud01}. None of these determinations are
problem-free: the $\nu_{26}$ band lacks spectroscopic data for the hot
band lines at 749.5~\cm , whereas the $\nu_{18}$ band suffers from a lack of
accurate continuum shape, and possible inaccuracies due to the use of
the JPL pseudo-linelist for the $\nu_6$ band of \ethane ,
for which no formal lists are available. Despite these uncertainties,
the values agree at a 2-$\sigma$ error level. We suspect that
systematic uncertainties (haze, ethane line list) are still of greater
concern for FP4 than FP3, and therefore we recommend that the lower
value $(4.2 \pm 0.5) \times 10^{-7}$, which agrees with previous analysis, is
considered valid at the 2 mbar level for low latitudes probed by these
retrievals. 

Further work is required to accurately characterize the broad
spectral features of Titan's mid-infrared continuum, as has been
performed for the far-infrared \citep{dekok07a}. 
Additional laboratory spectroscopy of alkanes is also required to
support modeling of planetary spectra. It is vital that high-quality
laboratory line lists are prepared for all bands of propane in the range
700--1300~\cm , and also the $\nu_6$ and $\nu_8$ bands of
ethane. The present lack of accurate spectral information is a serious
obstacle, not only to accurate measurement the abundance of propane on
Titan and the giant planets, and deriving underlying haze opacities, but also
because our inability to model and remove propane's prolific signature
in many infrared spectral regions is a major hindrance to the search
for the signatures of new stratospheric trace species.


\clearpage
{\bf Acknowledgements}

The acquisition of CIRS data is the result of the collective efforts of a
large number of people, including the following who worked on various
aspect of CIRS science planning, instrument commanding, uplink,
calibration and databasing: S.B. Calcutt, R.C. Carlson, M.H. Elliott,
E. Guandique, M. Kaelberer, E. Lellouch, A. Mamoutkine, P.J. Schinder,
M.E. Segura, J.S. Tingley, and also many engineers and science
planners at the Jet Propulsion Laboratory. We are indebted to
J. Vander Auwera, for supplying us with the recent ethane $\nu_9$ line
data base in advance of its inclusion in the next GEISA release, and
Steven Sharpe and colleagues at Pacific Northwest National Laboratory
(PNNL) for the laboratory propane absorption spectra. We thank Geoff
Toon and Armin Kleinboehl of JPL for the use of their  \ethane\
pseudo-linelist, which was generated using unpublished laboratory
spectra recorded with Fourier transform spectrometers at the PNNL and
the Kitt Peak National Observatory. We also thank W. Blass,
G.L. Bjoraker, S. Daunt and T. Fouchet for e-mail exchanges regarding
the GEISA propane data, and J.C. Brasunas for helpful discussions
regarding aliasing and numerical filtering in the instrument.
The US-based co-authors acknowledge the support of the NASA Cassini
Project during the period in which this work was completed.  

\appendix

\section{Spectral range of CIRS FP4}
\label{app:fp4}

The subject of the useable spectral range of CIRS FP4 is of
considerable interest in this paper, due to the existence of many
propane bands at the extremes of FP4 sensitivity. Further details of
the CIRS spectral bandpass are available in an internal CIRS team
report \cite{brasunas04}.

The nominal spectral range is 1100--1400~\cm , which
significantly underestimates the original bandpass: from
1020--1529~\cm , based on initial Nyquist sampling of the
interferogram signal in the instrument. Frequencies from outside this
range are aliased into the range, according to the formula:

\begin{equation}
\tilde{\nu}_{\rm nominal} +
\tilde{\nu}_{\rm aliased} = 2 \times
\tilde{\nu}_{\rm bandpass}
\end{equation}

A second-stage numerical filter is then applied that tapers the
response towards the ends of the range, and leads to a definition of
the `clean' bandpass where the strength of the aliased signal has a
relative response strength of less than 1\%:

\begin{equation}
\tilde{\nu}_{\rm clean}
= ( 2 \times \tilde{\nu}_{\rm bandpass}) -
\tilde{\nu}_{\rm 1pc}
\end{equation}

\noindent
where $\tilde{\nu}_{\rm 1pc}$ is defined as the (aliased) frequency
where the response of the numerical filter $R = F_{\rm FP4}(\tilde{\nu})$ has
dropped to 1\% of the peak response: $\tilde{\nu}_{\rm 1pc} = {F_{\rm
  FP4}}^{-1}(R_{\rm peak}\times 0.01)$. By this definition, the clean
bandpass of FP4 is 1131--1433~\cm , close to the canonical wavenumber
range (1100--1400 \cm ).

Between the Nyquist and `clean' bandpass limits at the low wavenumber
side (1020--1131 \cm ), the aliased signal comes from the region
889--1020~\cm , where there is little emission except from the
relatively weak lines of \ethylene . In addition, roll-off of detector
response at long wavelengths means that the `clean' bandpass is
somewhat underestimated. These effects mean that the range
1020--1131~\cm is very useful, especially as the aliased spectral
range 889--1020~\cm\ is known from the FP3 spectrum and its importance
can therefore be directly assessed.

The non-clean range from 1433--1529 is more problematic, because the
detector reponse remains strong, and more critically because we have
no  knowledge of the signal in the aliasing range above 1529~\cm ,
except from predictations of emission bands of known gases (e.g. the
weak $\nu_2$ band of methane at 1534 \cm ). Therefore, interpretation
of this spectral range must proceed with caution.

\section{Modifications to propane spectral line lists}

In this work we used propane line data from two sources: (i) the GEISA
atlas, for the $\nu_{26}$ band at 748~\cm , and (ii) recent
measurements by \citet{flaud01} for the bands in the range
1300--1500~\cm . During preliminary intercomparisons of the propane
abundance derived from the two spectral regions, we found large
differences in the retrieved values amounting to a factor $\sim 2$
(FP4 higher than FP3). We investigated possible causes of this
discrepancy, and discovered that the propane line strengths that are
in the GEISA 2003 atlas are incorrectly scaled: this problem is
expected to be fixed in the forthcoming GEISA 2009 release, that should
be available at the time of publication of this article. We also
suggest modifications to the Lorentzian Half-Width at Half Maximum
(HWHM) and temperature exponent of the spectral lines. A detailed
discussion follows.

\subsection{Radiative Transfer Treatment of Spectral Lines}

For atmospheric radiative transfer applications we wish to compute the
spectral dependence of the absorption co-efficient $k_\nu$ for each
molecular energy transition (i.e. spectral line). Absorption due to
all individual lines in a given spectral region is then included in
the overall opacity calculation, along with other sources
(particles, collision-induced opacity). Following \citet{goody89}, the
spectral absorption due to an individual line is:

\begin{equation}
k_\nu = Sf(\nu - \nu_0)
\end{equation}

\noindent
where $\nu$ is frequency, $\nu_0$ is the unperturbed frequency, $f$ is
a normalized function describing the line shape, and $S$ is the line
strength defined as:

\begin{equation}
S = \int k_\nu d\nu
\end{equation}

The lineshape function may be given by the classical Doppler (thermal)
or Lorentzian (pressure/collision) profiles, or more generally a
convolution of both, known as the Voigt profile:

\begin{equation}
f(\nu -\nu_0) = 
\int_{-\infty}^{+\infty} \frac{
\left(  \alpha_L / \pi \right) }
{
\left[ \left\{ (\nu- \nu_0) - (u\nu_0 / c) \right\}^2 +
{\alpha_L}^2 \right]
}
\left( \frac{m}{2\pi kT} \right) ^{\frac{1}{2}}
\exp \left ( \frac{-mu^2}{2kT} \right) du
\end{equation}

\noindent
In the above formula $k$, $h$ and $c$ are the usual constants of
Boltzmann, Planck and speed of light; $m$ is the molecular mass, $T$
is temperature and $\alpha_L$ is the Lorentzian HWHM. 

Therefore, in order to compute $k_\nu$ we must know the following:
$m$ and $T$, which are specified by the atmospheric problem; and
$\alpha_L$ and $S$, which are parameters obtained from experimental
measurement and tabulated in a database. Both of these parameters have
a temperature dependence, and in addition $\alpha_L$ is a function of
pressure also. First, the dependence of $\alpha_L$ on atmospheric
conditions is specified by:

\begin{equation}
\alpha_L (P,T) = \alpha_{L,0} \frac{P}{P_0} \left[ \frac{T_0}{T}
  \right]^n
\label{eq:lorentz}
\end{equation}

\noindent
where $(\alpha_{L,0},P_0,T_0)$ is a laboratory (usually) room
temperature measurement, and $n$ is a free scaling parameter known as the
Temperature Dependence of Width (TDW). Second, the line strength as a
function of temperature is given by \citep[e.g.][ p9033]{lacis91}:

\begin{equation}
S = S_0 \frac{V_0 R_0}{VR} \exp \left\{ \frac{hc}{k} E''
\left[ \frac{1}{T_0} - \frac{1}{T} \right] \right\}
\end{equation}

\noindent
where $E''$ is the energy of the lower state of the transition, and
$R$ and $V$ are the rotational and vibrational {\em partition functions}
(sums over microphysical state probabilities). The rotational
partition function is $R(T)=T^x$ where $x=1$ for linear molecules and
$x=3/2$ for non-linear molecules, such as propane. The vibrational
partition function is usually approximated by a quantum harmonic
oscillator so that:

\begin{equation}
V(T) = \prod_j \left( 1-\exp(-h\nu_j/kT) \right)^{-g_j}
\end{equation}

\noindent
where $\nu_j$ is the frequency of the $j^{th}$ vibrational mode, for
each of the 27 mode frequencies given in Table II, and $g_j$ is the
degeneracy of each mode, equal to 1 for this gas.

\subsection{GEISA Line Strengths for $\nu_{26}$}

The spectroscopic line list provided in the current GEISA 2003 edition
is unchanged since first added in 1991, and consists of $\sim 9000$
lines assigned by S.~Daunt (unpublished) from room temperature
spectra. S.~Daunt (private communication) has told us that the
integrated band intensity:

\begin{equation}
S = \int_i S_i d\nu
\end{equation}

\noindent
for $\sim3500$ lines for the fundamental mode only (excluding hotbands
- transitions from non-ground state energy levels) was scaled to match
the band sum for the entire region as reported by \citet{giver84},
which included hot bands. The value of $S$ reported by
\citet{giver84} is $4.33\times 10^{-19}$ \cmm , and we have separately
estimated from the PNNL propane spectra a value of $4.27\times
10^{-19}$ \cmm , in good agreement. However, the integrated band intensity of
the GEISA lines is slightly lower ($3.76\times 10^{-19}$ \cmm ).
We have calculated $V(296K) = 2.71$, which is approximately the ratio of
$S$ for all transitions to ground state transitions only, and
therefore, we estimate that the $S$ for the $\nu_{26}$
fundamental mode should be $(1/2.71) \times 4.27 \times 10^{-19} =
1.58 \times 10^{-19}$ \cmm . The strengths of individual lines in the atlas
should therefore by scaled by the ratio of the current GEISA $S$
to the predicted: i.e. by 1.58/3.76 = 0.420.
We have hence rescaled the GEISA lines for the $\nu_{26}$ band by this
factor prior to computation of the $k$-tables, which were used in our
retrievals.

\subsection{Lorentz Broadening Parameters}

In the current GEISA 2003 edition, the Lorentz HWHM $\alpha_L$ is
given as 0.08~\cm\ for the $\nu_{26}$ band. However, the FWHM for the
pressure broadening of \propane\ by \nitrogen\ has been separately
reported as 0.119~\cm\ by \citet{nadler89} (at 296 K) and 0.146~\cm\ by
\citet{hillman92} (at 175 K). The latter authors also used these two
measurements at different temperatures to infer a value for the $n$
exponent in Eq. \ref{eq:lorentz} of 0.50, which is lower than the 0.75
in GEISA at present.  We have therefore adopted values of
$\alpha_{L,0} = 0.12$ \cm\ and $n=0.50$ for all the bands of propane
available to us, including the FP4 bands.  

\clearpage

\bibliographystyle{harvard}


\def\baselinestretch{0.8}

\clearpage
\thispagestyle{empty}

\begin{table}[p]
\begin{centering}
\footnotesize
\begin{tabular}{cccccccc}
\multicolumn{8}{c}{\bf TABLE I} \\
\multicolumn{8}{c}{\bf Titan Observational Data} \\
\hline
Focal & Spectral & Latitude & Mean     & Altitude & Mean     & No. of
& NESR$^{\ddagger}$    \\ 
Plane & Range (\cm ) & Range    & Latitude & Range (km) & Alt.(km) 
& Spectra$^{\dagger}$ & (\radunit ) \\
\hline
FP3 & 600--1100 & 30\dg S--30\dg N & 9.5\dg N & 100--150 & 130 & 568 & $3\times
10^{-9}$ \\
FP4 & 1100-1500 & 30\dg S--30\dg N & 2.1\dg N & 100--150 & 125 & 616 & $3\times
10^{-10}$ \\
\hline
\end{tabular}
\normalsize
\newline
$^{\dagger}$ $2\times 10^{5}$ space and $2\times 10^{4}$ shutter 
spectra used in calibration.\\
$^{\ddagger}$ NESR = Noise Equivalent Spectral Radiance (1-$\sigma$). \\
\end{centering}
\label{tab:obs}
\end{table}

\clearpage
\thispagestyle{empty}

\begin{table}[p]
\begin{centering}
\footnotesize
\begin{tabular}{ccccc}
\multicolumn{5}{c}{\bf TABLE II} \\
\multicolumn{5}{c}{\bf Propane IR Spectral Bands} \\
\hline
Freq. & 1972 & 1965 &  Mode & Laboratory \\
(\cm ) & No.$^{\dagger}$ & No.$^{\ddagger}$ & 
Type & Spectra \\
\hline
 & & & & \\
\multicolumn{5}{c}{\bf -------------- CIRS FP1 --------------} \\
216 & 14 &  & Torsion & IR gas-phase inactive \\
268 & 27 &  & Torsion & \\
369 & 9 & 9 & CCC deform & \\
 & & & & \\
\multicolumn{5}{c}{\bf -------------- CIRS FP3 --------------} \\
748 & 26 & 21 & \methylene\ rock & Husson et al. (1992) \\
869 & 8 & 8 & CC stretch & \\
922 & 21 & 16 & \methyl\ rock &  \\
940 & 13 &  & \methyl\ rock & gas-phase inactive \\
 & & & & \\
\multicolumn{5}{c}{\bf ----------- CIRS FP3 \& FP4 -----------} \\
1053 & 20 & 15 & CC stretch & \\
 & & & & \\
\multicolumn{5}{c}{\bf -------------- CIRS FP4 --------------} \\
1158 & 7 & 7 & \methyl\ rock & \\
1192 & 25 & 20 & \methyl\ rock & \\ 
1278 & 12 &  & \methylene\ twist & gas-phase inactive \\
1338 & 19 & 14 & \methylene\ wag & Flaud et al. (2001) \\
1376 & 18 & 13 & \methyl\ s-deform & Flaud et al. (2001) \\
1392 & 6 & 6 & \methyl\ s-deform & \\
1451 & 11 &  & \methyl\ d-deform & gas-phase inactive \\
1462 & 5 & 5 & \methylene\ scissor & \\
1464 & 17 & 12 & \methyl\ d-deform & \\
1472 & 24 & 19 & \methyl\ d-deform & Flaud et al. (2001) \\
1476 & 4 & 4 & \methyl\ d-deform & Flaud et al. (2001) \\
 & & & & \\
\multicolumn{5}{c}{\bf -------------- CIRS ENDS --------------} \\
2887 & 16 & 3 & \methyl\ s-str & \\
2887 & 3 & 11 & \methylene\ s-str & \\
2962 & 2 & 2 & \methyl\ s-str & \\
2967 & 10 &  & \methyl\ d-str & gas-phase inactive \\
2968 & 23 & 18 & \methylene\ a-str & \\
2968 & 15 & 10 & \methyl\ d-str & \\
2973 & 22 & 17 & \methyl\ d-str & \\
2977 & 1 & 1 & \methyl\ d-str & \\
\hline
\end{tabular}
\normalsize
\newline
$^{\dagger}$Shimanouchi (1972)
$^{\ddagger}$Gayles and King (1965) \\
\end{centering}
\label{tab:bands}
\end{table}

\nocite{shimanouchi72, gayles65, husson92}

\clearpage
\thispagestyle{empty}

\begin{table}[p]
\begin{centering}
\begin{tabular}{ccc}
\multicolumn{3}{c}{\bf TABLE III} \\
\multicolumn{3}{c}{\bf Retrieved Propane VMR (ppm)} \\
\hline
 Band & Grey Haze & Tholin \\ 
\hline
 748 \cm\ & (4.2$\pm$0.5) $\times 10^{-7}$ & (4.3$\pm$0.5) $\times 10^{-7}$ \\ 
1376 \cm\ & (5.7$\pm$0.8) $\times 10^{-7}$ & (5.7$\pm$0.8) $\times 10^{-7}$ \\ 
\hline

\end{tabular}
\end{centering}
\label{tab:rets}
\end{table}

\def\baselinestretch{1.6}

\clearpage

{\bf Figure Captions}

Fig. \ref{fig:gas_apriori} 
Initial gas profiles for Titan spectral model. Upper plot:
major isotopic species. Lower plot: minor isotopic species.

Fig. \ref{fig:limb_convs}
Plots of $W(z_i)$ spatial weighting function for limb
spectral averages, resampled onto a 10~km grid relative to the mean
tangent altitude, and normalized such that $\Sigma_i W(z_i) = 1.0$. 
Dashed line: FP3. Solid line: FP4.  The effective weighting function
of the final mean FP3 and FP4 spectra is different, due to two
effects: (i) the different real detector spatial responses, and (ii)
the slightly different distributions of center altitudes of the
spectral sets comprising each average. 

Fig. \ref{fig:ret_temps}
A priori (dashed line) and retrieved temperature profile
(solid line, with 1$\sigma$ error bars) from Titan low-latitude (30\dg
S--30\dg N) spectral average from limb data (100--150 km tangent
altitude). The grey shaded box shows the approximate region of
real temperature information in the retrieval, as bounded by the min
and max values of FWHM pressure levels of all temperature contribution
functions assessed across the entire wavenumber range. At higher and
lower levels the temperature profiles is smoothly joined to the {\em a
  priori}. A second weaker information region also exists above
$\sim$500 km due to methane hotband emissions.

Fig. \ref{fig:terr}
Comparison of various contributions to the error on retrieved
temperatures from fitting the methane $\nu_4$ band. Dotted line:
initial \apriori\ error estimate. Dashed line: retrieval error due to
spectral noise. Dot-dashed line: error due to methane abundance
profile uncertainty. The combination of these two (spectral and
methane profile errors) lead to the final vertical temperature error:
solid line.

Fig. \ref{fig:fp3_propane}
Identification of propane spectral bands from CIRS Focal
Plane 3 (600--1100~\cm ). Left column: CIRS averaged limb spectra (9\dg N,
130~km), four selected data ranges (black line) and synthetic spectral
fit (red) with all  gases except \propane\ in model. Right column: residual
spectrum after model subtraction (red) compared to PNNL laboratory
absorption spectra (blue) and 1-$\sigma$ NESR (grey). Propane bands
are identified by text and dashed lines. Note: incomplete removal of
\ethane\ $\nu_9$ band emission is evident in the residual (d)at 840--850~\cm .

Fig. \ref{fig:fp4_propane}
Identification of propane spectral bands from CIRS Focal
Plane 4 (1000--1500~\cm ). Left column: CIRS averaged limb spectra (2\dg N,
125~km), four selected data ranges (black line) and synthetic spectral
fit (red) with all  gases exept \propane\ in model. Right column: residual
spectrum after model subtraction (red) compared to PNNL laboratory
absorption spectra (blue) and 1-$\sigma$ NESR (grey). Propane bands
are identified by text and dashed lines. Note: spectral baseline in
the range 1440--1480~\cm\ (panel (h)) is not well modeled by grey
haze, leading to  over-estimation of the ethane abundance (features
marked with `*') as the retrieval attempts to minimize the overall
$\chi^2$ of the fit.

Fig. \ref{fig:prop_rets_rescale}
Residuals of fitting the two propane spectral regions. 
Red lines: residual of forward calculation with no propane in model
atmosphere, all other gases at previously retrieved values. Blue line:
residual after fitting with propane included, and grey haze
model. Green line: residual after fitting with propane included, and
tholin particle haze model. Grey bands are the 1-$\sigma$ NESR. 
(a) $\nu_{26}$ band region, fitted by GEISA 2003 propane line
data modified as described in Appendix B. Note that the propane
spectral data does not include the hot band at 749.5 \cm , and also the
significant difference to the `baseline' level due to propane. (b) The
$\nu_{18}$ band fitted with the recent spectral data of Flaud et al. (2001).


\LARGE

\clearpage

\begin{figure}[t]\centerline{
\epsfig{file=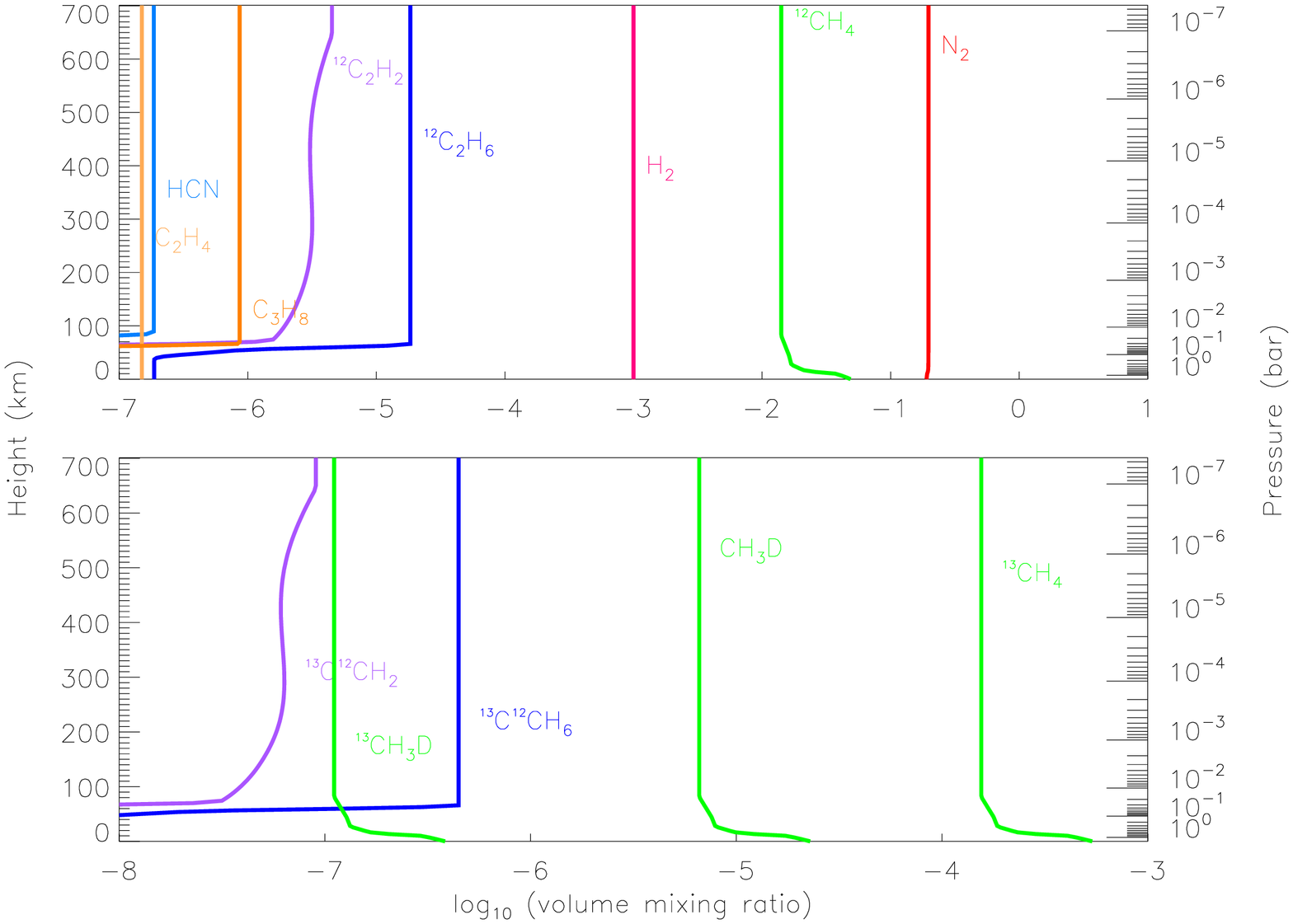, height=19cm, angle=90}
}
\end{figure}

Nixon et al. {\bf Figure \ref{fig:gas_apriori}}

\clearpage

\begin{figure}[t]\centerline{
\epsfig{file=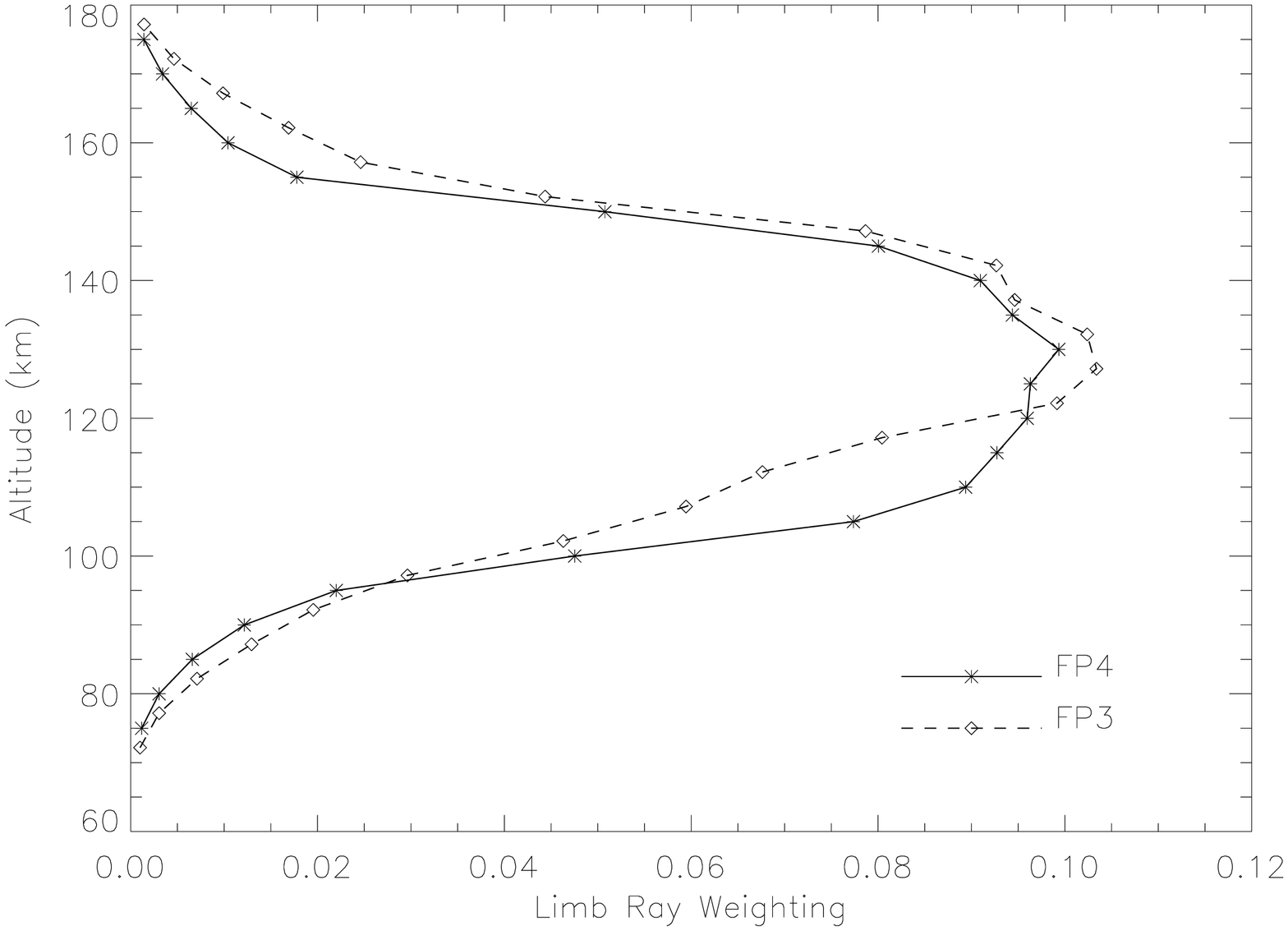, height=19cm, angle=90}
}
\end{figure}

Nixon et al. {\bf Figure \ref{fig:limb_convs}}

\clearpage

\begin{figure}[t]\centerline{
\epsfig{file=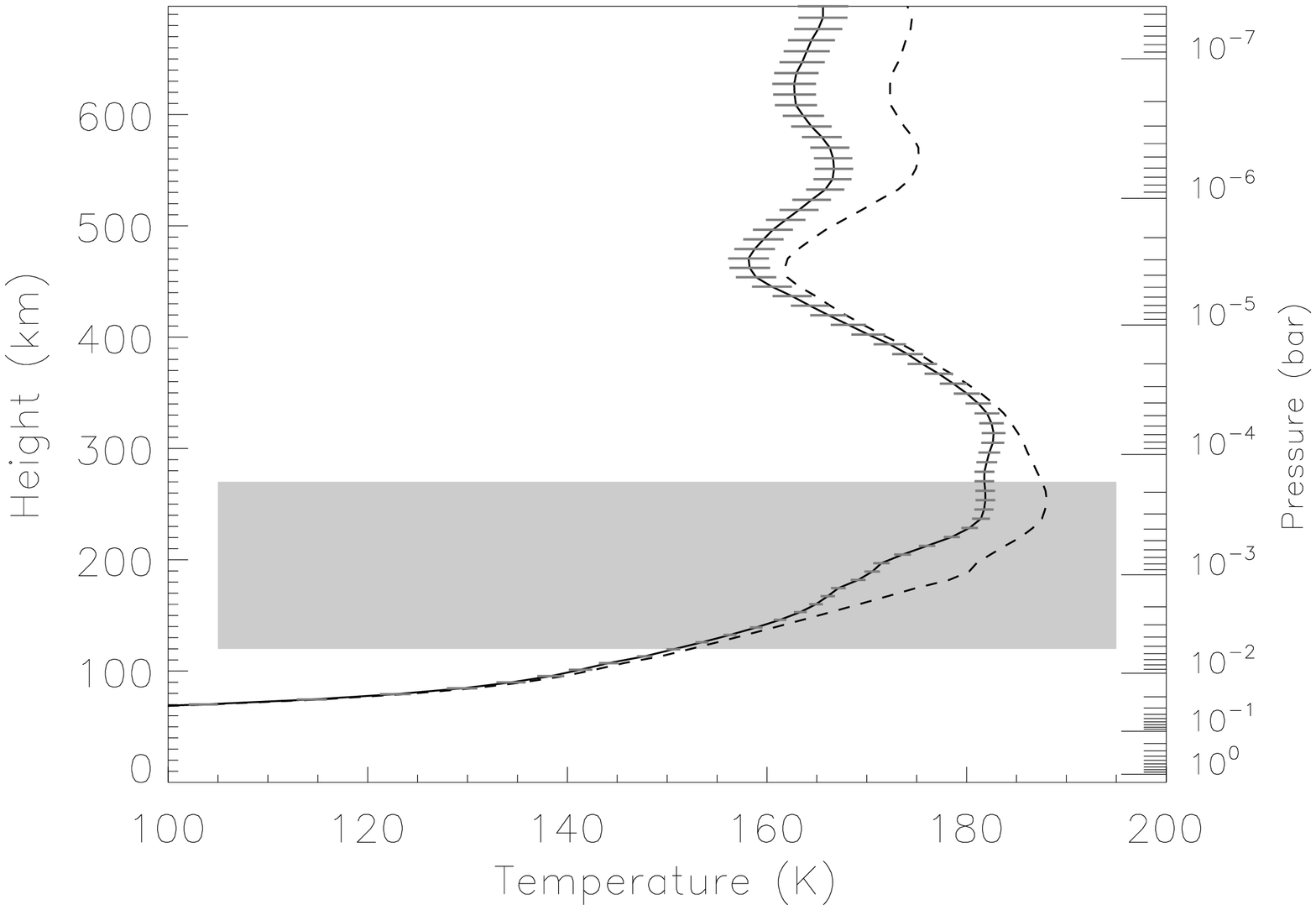, height=19cm, angle=90}
}
\end{figure}

Nixon et al. {\bf Figure \ref{fig:ret_temps}}

\clearpage

\begin{figure}[t]\centerline{
\epsfig{file=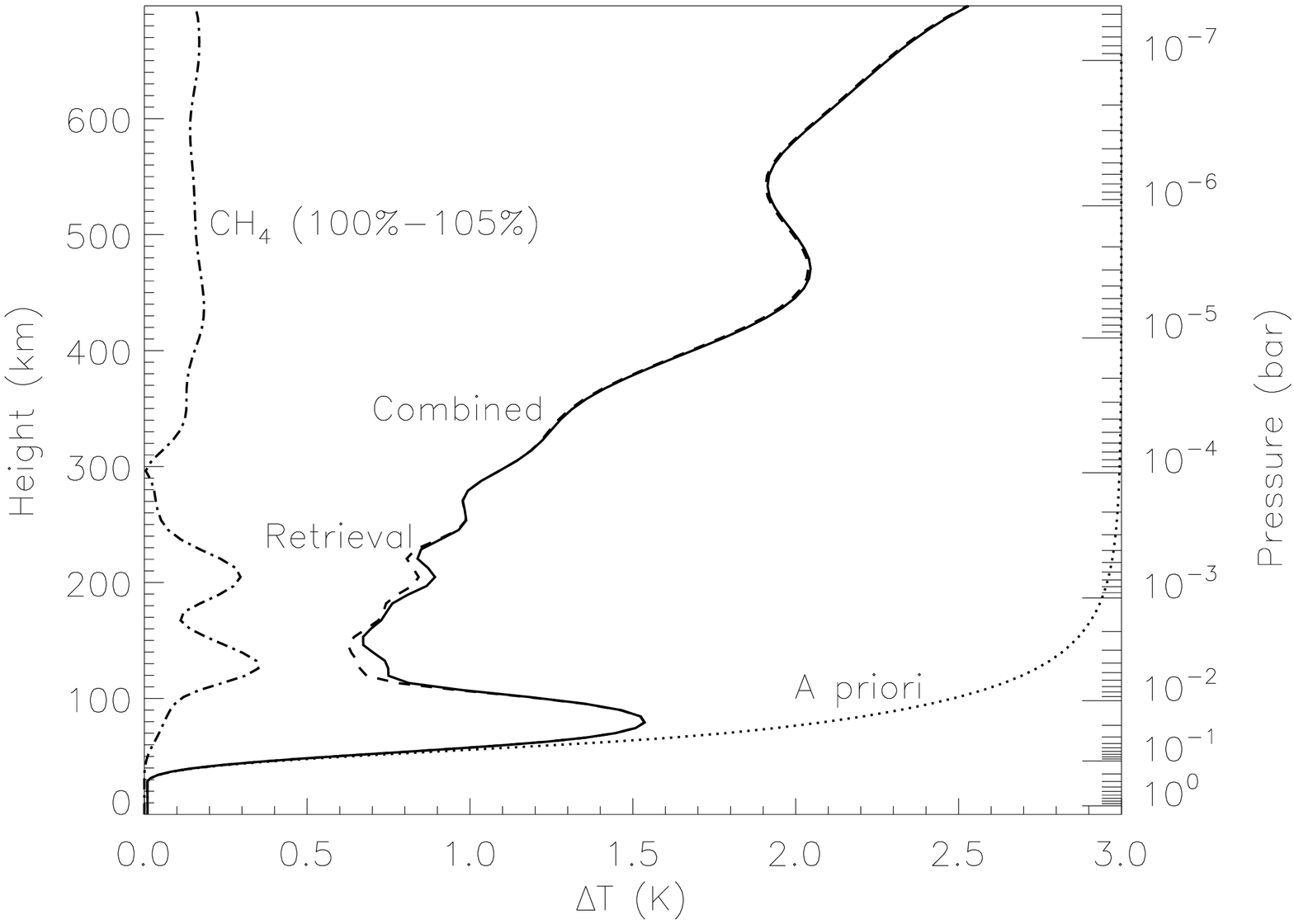, height=19cm, angle=90}
}
\end{figure}

Nixon et al. {\bf Figure \ref{fig:terr}}

\clearpage

\begin{figure}[t]\centerline{
\epsfig{file=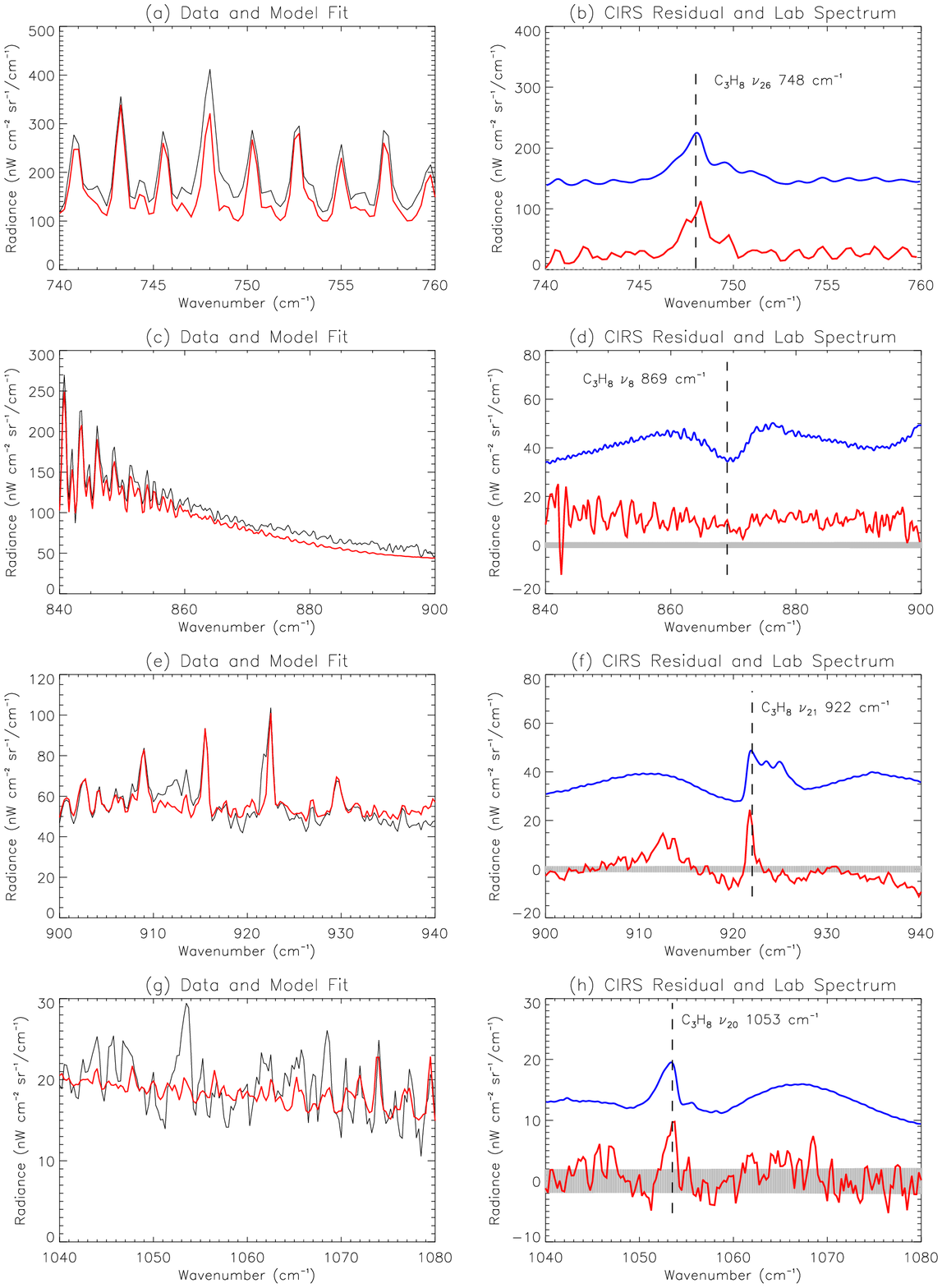, height=15cm, angle=0}
}
\end{figure}

Nixon et al. {\bf Figure \ref{fig:fp3_propane}}

\clearpage

\begin{figure}[t]\centerline{
\epsfig{file=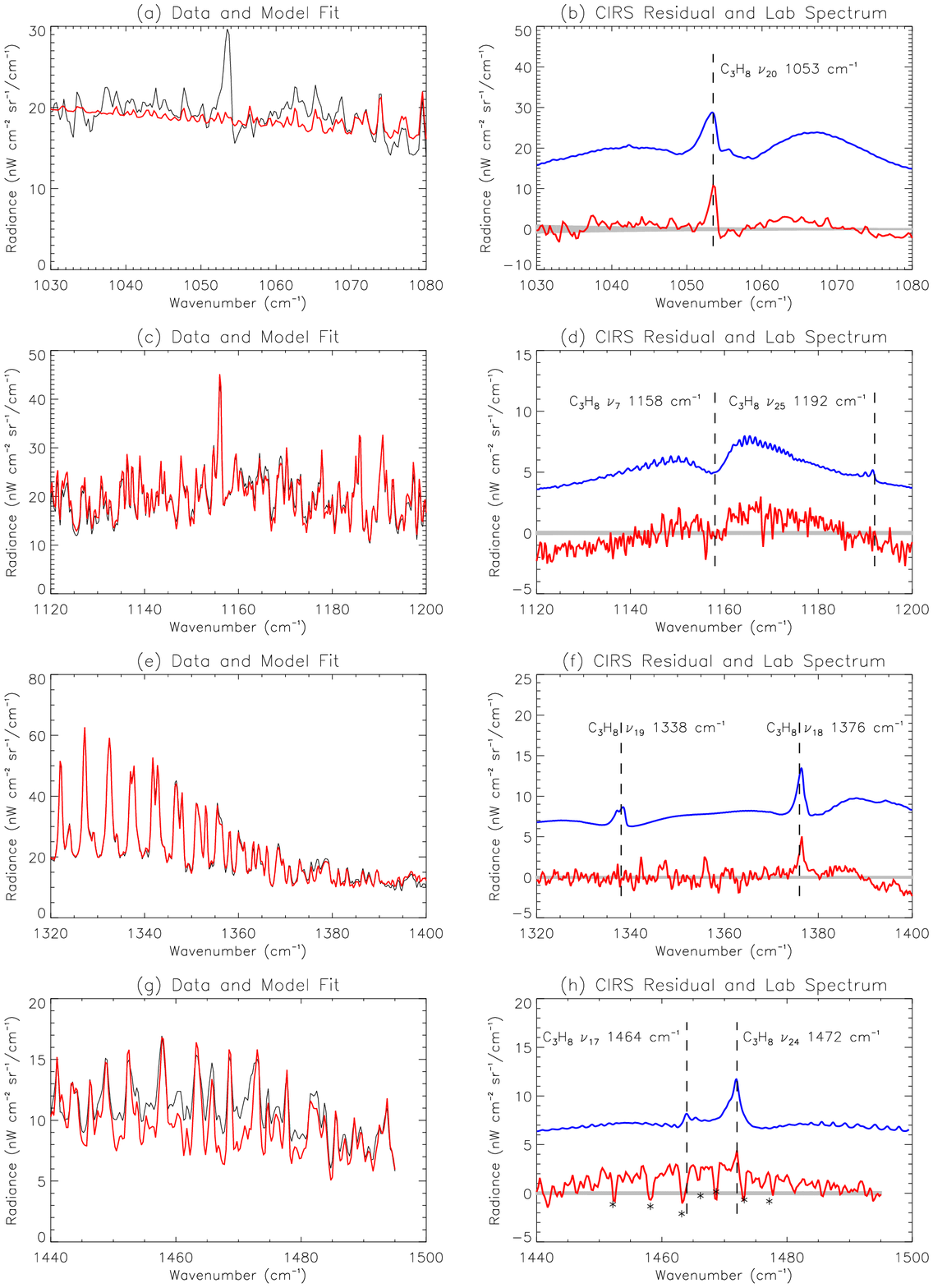, height=15cm, angle=0}
}
\end{figure}

Nixon et al. {\bf Figure \ref{fig:fp4_propane}}

\clearpage

\begin{figure}[t]\centerline{
\epsfig{file=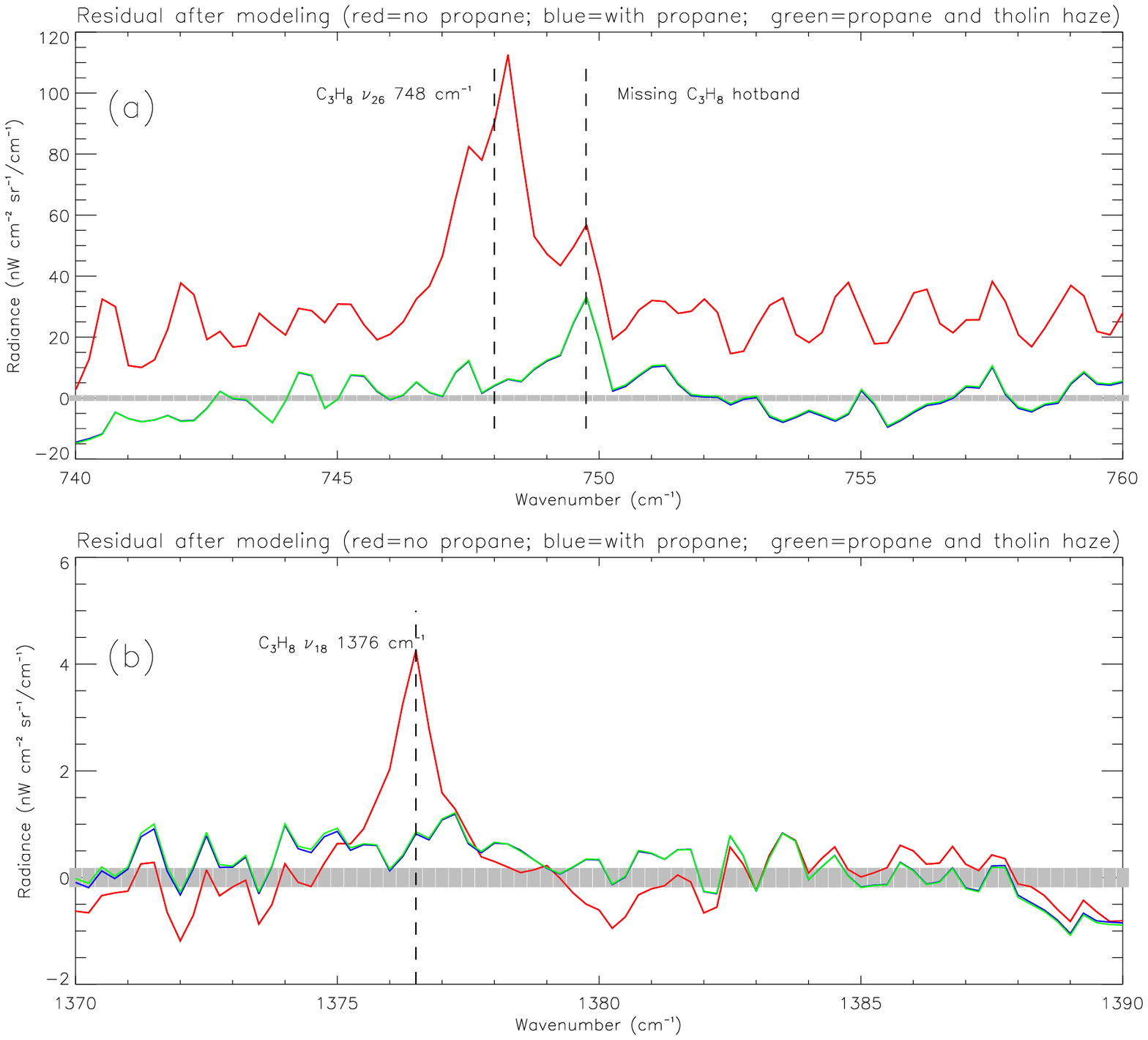, height=15cm, angle=0}
}
\end{figure}

Nixon et al. {\bf Figure \ref{fig:prop_rets_rescale}}

\end{document}